\documentclass[10pt,conference]{IEEEtran}

\usepackage{epsfig,amsfonts,amsmath,amssymb,verbatim}
\usepackage[usenames]{color}

\def\eE{{\mathbb E}}

\newcommand{\secref}[1]{Section~\ref{#1}}
\newcommand{\eqnref}[1]{(\ref{#1})}
\newcommand{\thmref}[1]{Theorem~\ref{#1}}

\newcommand{\figref}[1]{Figure~\ref{#1}}

\def\@begintheorem#1#2{\tmpitemindent\itemindent\topsep 0pt\rm\trivlist
    \item[\hskip \labelsep{\indent\it #1\ #2:}]\itemindent\tmpitemindent}
\def\@opargbegintheorem#1#2#3{\tmpitemindent\itemindent\topsep 0pt\rm \trivlist
    \item[\hskip\labelsep{\indent\it #1\ #2\
    \rm(#3):}]\itemindent\tmpitemindent}
\def\@endtheorem{\endtrivlist\unskip}

\newtheorem{theorem}{Theorem}[section]

\newtheorem{corollary}{Corollary}[section]

\begin{document}

\title{Bootstrapped Oblivious Transfer and Secure Two-Party Function Computation$^{\text{\small 1}}$}

\author{
\authorblockN{Ye Wang and Prakash Ishwar}
\authorblockA{Department of Electrical and Computer Engineering\\
Boston University \\
Boston, MA \\
Email: \{yw,pi\}@bu.edu}}

\maketitle

\begin{abstract}
We propose an information theoretic framework for the secure
two-party function computation (SFC) problem and introduce the
notion of SFC capacity. We study and extend string oblivious
transfer (OT) to {\em sample-wise} OT. We propose an efficient, {\em
perfectly private} OT protocol utilizing the binary erasure channel
or source. We also propose the {\em bootstrap} string OT protocol
which provides disjoint (weakened) privacy while achieving a
multiplicative increase in rate, thus trading off security for rate.
Finally, leveraging our OT protocol, we construct a protocol for SFC
and establish a general lower bound on SFC capacity of the binary
erasure channel and source.
\end{abstract}


\section{Introduction}

\addtocounter{footnote}{+1} \footnotetext{This material is based
upon work supported by the US National Science Foundation (NSF)
under award (CAREER) CCF--0546598. Any opinions, findings, and
conclusions or recommendations expressed in this material are those
of the authors and do not necessarily reflect the views of the NSF.}

Motivated by applications ranging from confidential database access
to oblivious contract negotiation \cite{Yao-MPC82}, we study the
problem of {\em secure two-party function computation} (SFC). In
this problem, Alice and Bob each have private data, and they wish to
compute functions of both of their data. The objective is to design
a protocol that ensures {\em correctness} of the computed functions
while maintaining individual {\em privacy}, in the sense that
neither party gains any information about the other's data other
than what can be inferred from the result of their function
computation. An important special case of this problem is {\em
string oblivious transfer} (OT) from \cite{EvenGL-ACM85-OTPaper},
wherein Alice has two strings $\tilde{A}_0$ and $\tilde{A}_1$ and
Bob has a single bit $B$. An OT protocol should reveal $\tilde{A}_B$
to Bob, while Alice remains ignorant of $B$ and Bob of
$\tilde{A}_{(1-B)}$.

In this work, we propose an information theoretic framework for SFC
and introduce the notion of SFC rates and capacity, in terms of the
ratio of samples of computation to samples of {\em correlated
randomness} needed. Correlated randomness is a noisy resource in the
form of a noisy communication channel or distributed random source
available between the parties. We cast the string OT problem as a
special case within our framework and also introduce the {\em
sample-wise} OT problem. We address the string and sample-wise OT
problems with an efficient {\em perfectly private} protocol
utilizing the binary erasure channel or source. For the string OT
problem, we also propose the {\em bootstrap} protocol which provides
disjoint (weakened) privacy while achieving a multiplicative
increase in rate, thus trading off security for rate. Finally,
leveraging our OT protocol, we construct a protocol for SFC and
establish a general lower bound on SFC capacity binary erasure
channel and source. Due to space limitations, detailed proofs are
omitted, but will appear in an extended version of this work.

Our objective is information theoretic (unconditional) security,
where even computationally unbounded adversaries must not be able to
break the privacy. We work with the assumption of {\em semi-honest}
(or passive) parties, where the parties honestly follow the
protocol. It is well-known that in this setting both OT and SFC
cannot be realized ``from scratch'' \cite{Kilian-ACM88-CryptoFromOT,
CramerD-2005-MPCaI}, that is with protocols using only noise-free
communication channels and local randomness. It has been observed
that OT becomes possible given correlated randomness
\cite{Crepeau-EuroCrypt-97-CrypBasedOnNoisyChans,
CrepeauK-FOCS-88-OTusingWeakSecure,
NascimentoW-ISIT06-OblivTransNoisyCorr,
AhlswedeC-ISIT07-OblivTransCap}, and that SFC also becomes possible
based on OT \cite{Kilian-ACM88-CryptoFromOT}. Thus, correlated
randomness is a valuable resource as an enabling factor for OT and
SFC. Recently the concept of {\em OT capacity} of a channel or
source, measuring the fundamental limit of how efficiently the
resource can be used toward OT, has been introduced in
\cite{NascimentoEtAl-ISIT06-OblivTransErasChan,
NascimentoW-ISIT06-OblivTransNoisyCorr} and further characterized by
\cite{AhlswedeC-ISIT07-OblivTransCap,
NascimentoW-IT2008-OTCapOfNoisyRes}.


\section{Problem Formulation} \label{sec:Problem}

In this section, we first formulate the SFC problem within a novel
information theoretic framework. This framework utilizes conditional
mutual information based privacy measures, and defines achievable
function computation rate and capacity. We then discuss OT which is
encompassed by this SFC framework as a special case. In the last
subsection we present the problem of string OT with the notion of
{\em disjoint privacy}.

\subsection{Secure Two-Party Function Computation} \label{sec:probS2PFC}
Two parties, Alice and Bob, each have $k$ samples of a jointly
distributed source on the finite alphabets $\mathcal{A} \times
\mathcal{B}$, where Alice possesses $A^k \triangleq \{A_1, \ldots,
A_k\} \in \mathcal{A}^k$ and Bob possesses $B^k \triangleq \{B_1,
\ldots, B_k\} \in \mathcal{B}^k$, with $(A^k,B^k) \sim P_{A^k,B^k}$.
For a given function $f:\mathcal{A} \times \mathcal{B} \rightarrow
\mathcal{R}_f$, Alice wishes to compute samples of a function of the
sources $F^k \triangleq \{f(A_1,B_1), \ldots, f(A_k,B_k)\}$.
Similarly, Bob wishes to compute $G^k \triangleq \{g(A_1,B_1),
\ldots, g(A_k,B_k)\}$ where $g:\mathcal{A} \times \mathcal{B}
\rightarrow \mathcal{R}_g$. Alice and Bob cooperatively compute
these functions via an interactive protocol that may exchange
messages over an error-free discussion channel and also utilize $n$
samples of {\em correlated randomness}. The correlated randomness is
a precious resource which comes in two possible forms:
\begin{itemize}
\item {\em Source-model:} For $i = 1,\ldots,n$, $(X_i, Y_i)
\stackrel{\mathrm{iid}}{\sim} P_{X,Y}$. $X^n \triangleq (X_1,
\ldots, X_n)$ is available to Alice and $Y^n \triangleq (Y_1,
\ldots, Y_n)$ to Bob.
\item {\em Channel-model:} $X^n$ and $Y^n$ are respectively the
sequence of inputs and outputs of a discrete memoryless channel
(DMC) with conditional distribution $P_{Y|X}$, with $X^n$ selected
by Alice and $Y^n$ received by Bob.
\end{itemize}

An acceptable $(n,k)$-protocol for {\em source-model} correlated
randomness is defined as follows. First, Alice and Bob receive
$(A^k,X^n)$ and $(B^k,Y^n)$, and generate local random variables
$Z_A$ and $Z_B$ respectively, where $Z_A$, $Z_B$, $(A^k,B^k)$, and
$(X^n,Y^n)$ are mutually independent. Then, over $r$ stages, Alice
and Bob exchange messages $M_1, \ldots, M_r$ over the error-free
discussion channel, where in an odd numbered stage $i$ Alice
produces message $M_i$ as a function of everything available to her,
namely $(A^k,X^n,Z_A,M^{i-1})$, and in an even numbered stage $j$
Bob produces the messages $M_j$ as a function of everything
available to him, namely $(B^k,Y^n,Z_B,M^{i-1})$. At the end of the
protocol, Alice and Bob produce function estimates $\hat{F}^k$ and
$\hat{G}^k$ as functions of $(A^k,X^n,Z_A,M^r)$ and
$(B^k,Y^n,Z_B,M^r)$ respectively.

An acceptable $(n,k)$-protocol for {\em channel-model} correlated
randomness is similar to the source-model protocol, but $(X^n,Y^n)$
are not given at the beginning of the protocol. Instead, the samples
$X^n$ are generated by Alice, transmitted into the DMC, and outputs
$Y^n$ are received by Bob. The DMC transmissions may be arbitrarily
interspersed with discussion stages (including happening entirely
before or after the discussion messages are exchanged). At each
stage or transmission, the discussion message or channel input
symbol is a function of everything available to the sending party. A
source-model protocol can be realized as a special case of the
channel-model if Alice randomizes the inputs, for $i = 1,\ldots,n$,
$X_i \stackrel{\mathrm{iid}}{\sim} P_X$, and transmits before any
discussion messages are sent.

For both models, $R > 0$ is a called an {\em achievable SFC rate}
for the particular sources, functions, and correlated randomness if
for every $\epsilon
> 0$, and all sufficiently large $n$, there exists an acceptable
$(n,k)$-protocol with $(k/n) > R - \epsilon$ satisfying the
following
\begin{itemize}
\item {\em (Correctness)} $\Pr[\hat{F}^k \neq F^k] < \epsilon$ and
$\Pr[\hat{G}^k \neq G^k] < \epsilon$,
\item {\em (Privacy for Alice)} \begin{eqnarray}
\label{eqn:alicePriv} I(A^k;Z_B,Y^n,M^r|B^k,G^k) < \epsilon,
\end{eqnarray}
\item {\em (Privacy for Bob)} \begin{eqnarray}
\label{eqn:bobPriv} I(B^k;Z_A,X^n,M^r|A^k,F^k) < \epsilon.
\end{eqnarray}
\end{itemize}
A protocol is said to be {\em perfectly private} if the privacy
constraints of \eqnref{eqn:alicePriv} and \eqnref{eqn:bobPriv} are
exactly zero. The {\em SFC capacity} $C$ for the particular sources,
functions, and correlated randomness is defined as the largest
achievable function computation rate, and $0$ if no rate $R > 0$ is
achievable.

\subsection{Sample-wise Oblivious Transfer} \label{sec:probSWOT}
The $1$-out-of-$m$ sample-wise OT problem is a special case of the
SFC problem, wherein Alice's source alphabet is $\mathcal{A} =
\{0,1\}^m$, Bob's source alphabet is $\mathcal{B} = \{1,\ldots,m\}$,
Alice's function is constant $f = 0$, and Bob's function is given by
$g((a_1,\ldots,a_m),b) = a_b$. For clarity of exposition, let
$\mathbf{A}$ be the $k \times m$ binary matrix formed by vertically
stacking Alice's $m$-bit samples $A_1, \ldots, A_k$ as the rows. Bob
wishes to receive the $k$ bits $G_1, \ldots, G_k$, where $G_i =
\mathbf{A}_{i,B_i}$. Alice's privacy condition
\eqnref{eqn:alicePriv} means that Bob obtains no information about
the other $k(m-1)$ bits of $\mathbf{A}$ that he did not select.
Bob's privacy condition \eqnref{eqn:bobPriv} means that Alice
obtains no information about Bob's selection $B^k$. When dealing
with the above scenario, we speak of achievable sample-wise OT rate
$R_{OT,m}$, and the sample-wise OT capacity $C_{OT,m}$.

\subsection{String Oblivious Transfer with Disjoint Privacy} \label{sec:probWeakOT}
The $1$-out-of-$m$ string OT problem is a special case of the
$1$-out-of-$m$ sample-wise OT problem, wherein the source
distribution is specified as
\[
P_{A^k,B^k}(a^k,b^k) = \begin{cases}
\frac{1}{m 2^{km}}, & \mbox{if } b_1 = \ldots = b_k \\
0, & \mbox{otherwise,}
\end{cases}
\]
that is, Alice's source samples $A^k$ consist of $km$ iid
Bernoulli-$(1/2)$ bits and is independent of Bob's source $B^k$
which always consists of identical samples uniformly distributed
over $\mathcal{B} = \{1,\ldots,m\}$. Interpreting this scenario,
Alice has $m$, $k$-bit strings $\tilde{A}_1, \ldots, \tilde{A}_m$,
which are aligned as the {\em columns} of $\mathbf{A}$, and Bob has
the selection $B \triangleq B_1$ and wishes to receive the $k$-bit
string $\tilde{A}_B$.

Alice's privacy condition \eqnref{eqn:alicePriv} reduces to
\begin{eqnarray*}
I(\{\tilde{A}_i\}_{i=1,i \neq B}^{k};Z_B,Y^n,M^r|B,\tilde{A}_B) <
\epsilon,
\end{eqnarray*}
which implies that Bob is unable to reconstruct any string that he
did not select or any non-trivial {\em joint function} of the
strings that he did not select without non-negligible probability of
error. The interesting alternative notion of {\em disjoint privacy}
replaces Alice's privacy condition \eqnref{eqn:alicePriv} with
\begin{equation} \label{eqn:weakPrivacy}
\mbox{for } i \in \{1,\ldots,m\},\quad
I(\tilde{A}_i;Z_B,Y^n,M^r|B,\tilde{A}_B) < \epsilon,
\end{equation}
which implies that Bob is unable to reconstruct any non-trivial
function of any individual string (including the string itself) that
he did not select without non-negligible probability of error. A
protocol that satisfies \eqnref{eqn:alicePriv} will also satisfy
this disjoint privacy condition \eqnref{eqn:weakPrivacy}, however
the converse is not true. For example, a protocol that reveals to
Bob $\tilde{A}_B$ and also the binary exclusive-or (XOR) all of the
strings $\tilde{A}_1 \oplus \ldots \oplus \tilde{A}_m$ will satisfy
the disjoint privacy constraint, but will not satisfy
\eqnref{eqn:alicePriv}. A protocol obtains {\em perfect disjoint
privacy} if the privacy constraints of \eqnref{eqn:weakPrivacy} are
exactly zero.

The motivation for considering the weakened sense of disjoint
privacy is to explore protocols (see \secref{sec:Boot}) that
tradeoff privacy in order to achieve higher rates than the
sample-wise OT protocol of \secref{sec:protocolSWOT}. When dealing
with the above scenario with the disjoint privacy condition
\eqnref{eqn:weakPrivacy} replacing Alice's standard privacy
condition \eqnref{eqn:alicePriv}, we speak of achievable string OT
rate with disjoint privacy $\widetilde{R}_{OT,m}$, and string OT
capacity with disjoint privacy $\widetilde{C}_{OT,m}$.


\section{Sample-wise Oblivious Transfer}
In this section, we present the sample-wise oblivious transfer ({\bf
SWOT}) protocol. Later on, by leveraging the {\bf SWOT} protocol, we
construct protocols for string OT with disjoint privacy (see
\secref{sec:protocolBootOT}) and for SFC (see
\secref{sec:protocolFuncComp}). The {\bf SWOT} protocol utilizes
correlated randomness in form of the binary erasure channel (BEC)
and the binary erasure source (BES). The BEC$(p)$ has the input and
output alphabets $\mathcal{X} = \{0,1\}$ and $\mathcal{Y} =
\{0,1,e\}$, with the conditional distribution $P_{Y|X}(y|x) =
p\mathbf{1}_{(y=e)} + (1-p)\mathbf{1}_{(y=x)}$, where $p$ is the the
{\em probability of erasure}. The BES$(p)$ has the joint
distribution $P_{X,Y}(x,y) = (1/2)P_{Y|X}(y|x)$. The protocol will
be described as a source-model protocol for the BES$(p)$, which is
easily adapted into a channel-model protocol for the BEC$(p)$ by
adding the initial step of Alice transmitting $n$ iid
Bernoulli-$(1/2)$ bits into the BEC in order to simulate $n$ samples
of a BES. Because of the interchangeability of the BEC and BES, we
will write BES/BEC$(p)$ to denote that the correlated randomness is
either the BEC$(p)$ or BES$(p)$.

\subsection{Sample-wise Oblivious Transfer Protocol} \label{sec:protocolSWOT}
This protocol is inspired by the protocols for the binary erasure
channel given by \cite{AhlswedeC-ISIT07-OblivTransCap,
NascimentoEtAl-ISIT06-OblivTransErasChan}. The novel aspects of our
protocol are the treatment of {\em sample-wise} as opposed to {\em
string} oblivious transfer and the mechanism of failing into an
error case when privacy cannot be provided. This yields a perfectly
private protocol with roughly the same negligible probability of
error, and also simplifies analysis of both privacy and error. The
perfect privacy of this protocol is important since it enables it to
be leveraged in a secure black-box manner to construct other
protocols without complicating the analysis of privacy. The basic
idea of this protocol is to use the erasures of the BES/BEC to
conceal the $k(m-1)$ bits at the locations in $\mathbf{A}$ that Bob
must remain ignorant of, while using the non-erasures to reveal the
$k$ bits at the locations in $\mathbf{A}$ that Bob has selected.

\begin{itemize}
\item Bob partitions $\{1, \ldots, n\}$ into the set of locations of
erasures $S_e$ and locations of non-erasures $S$ in $Y^n$, that is,
$Y_i = e$ if and only if $i \in S_e$, and $S = \{1, \ldots, n\}
\setminus S_e$.
\item If there is not enough erasures or non-erasures, $k > |S|$ or
$k(m-1) > |S_e|$, then the protocol aborts and Bob sets his function
estimate to $\hat{G}^k = 0^k$.
\item Otherwise, the protocol continues and Bob creates an $k \times m$
matrix $\mathbf{U}$, where at each of the $k$ positions specified by
$\{(i,B_i):i = 1,\ldots,k\}$, a random, uniform selection, without
replacement, from $S$ is placed. Similarly, at the other $k(m-1)$
positions, a random selection from $S_e$ is placed. Bob sends the
matrix $\mathbf{U}$ to Alice via the discussion channel.
\item The $k \times m$ matrix $\mathbf{U}$, whose elements belong to
$\{1,\ldots,n\}$ specifies how Alice should arrange $km$ of the $n$
bits $X^n$ into the $k \times m$ binary matrix $\mathbf{X_U}$, via
$\mathbf{X_U}(i,j) = X_{\mathbf{U}(i,j)}$. Alice computes
$\mathbf{C} = \mathbf{A} \bigoplus \mathbf{X_U}$, where $\bigoplus$
denotes element-wise binary exclusive-or (XOR), and sends
$\mathbf{C}$ to Bob over the discussion channel.
\item Bob is able create his function estimate $\hat{G}^k$ by
reversing the XOR since $\mathbf{Y_U}$ is equal to $\mathbf{X_U}$ at
locations corresponding to the locations of $\mathbf{A}$ that he has
selected.
\end{itemize}

\subsection{Analysis and Achievable Rates} \label{sec:mainSWOT}
The {\bf SWOT} protocol is perfectly private for Bob since any
$\mathbf{U}$ is uniformly possible given any realization of Bob's
samples $B^k$ because erasures uniformly and independently occur in
$Y^n$. The protocol is perfectly private for Alice since
$\mathbf{C}$ is only sent to Bob if the protocol does not abort and
there have been enough erased bits, acting as a Bernoulli-$(1/2)$
one-time pad, to mask the $k(m-1)$ bits that should be concealed.
The protocol is correct if it does not abort, thus the probability
of error is bounded by the probability of aborting, which becomes
negligible for $n$ sufficiently large if $ k < n(1-p) = \eE|S|$ and
$k(m-1) < np = \eE|S_e|$, by the law of large numbers. Thus, the
rate $R_{OT,m} = \min((1-p),p/(m-1))$ is achievable by this
protocol. This protocol is distribution-free since the above
arguments hold not only for any distribution, but also for any
realization of the sources $(A^k,B^k)$. These results are summarized
in the following theorem

\begin{theorem} \label{thm:SWOTRateLower}
For any arbitrary source distribution $P_{A_k,B_k}$, the {\bf SWOT}
protocol, utilizing correlated randomness in the form a
BES/BEC$(p)$, obtains perfect privacy and achieves the
$1$-out-of-$m$ sample-wise OT rate
\[
R_{OT,m} = \min((1-p),p/(m-1)).
\]
Hence, the $1$-out-of-$m$ sample-wise OT capacity for a BES/BEC$(p)$
and for arbitrary source distributions is bounded below by
\[
C_{OT,m} \geq R_{OT,m} = \min((1-p),p/(m-1)).
\]
\end{theorem}

An upper bound to the sample-wise OT capacity for general correlated
randomness is established by the following theorem.

\begin{theorem} \label{thm:SWOTCapUpper}
For the uniform source distribution and general source-model
correlated randomness, we have
\[
C_{OT,m} \leq \min(I(X;Y),H(X|Y)/(m-1)).
\]
For channel-model correlated randomness, the right side of the above
expression is maximized over $P_{X}$.
\end{theorem}

The proof of this theorem is omitted due to space limitations. It
follows from the methods and results used in
\cite[Theorem~1]{AhlswedeC-ISIT07-OblivTransCap} and
\cite[Lemma~7]{NascimentoEtAl-ISIT06-OblivTransErasChan}.

For the BES$(p)$, $H(X|Y) = p$ and $I(X;Y) = 1-p$. For the BEC$(p)$,
$\max_{P_X} \min(I(X;Y),H(X|Y)/(m-1)) = \min((1-p),p/(m-1))$ with
the maximum achieved by $P_X = (1/2)$. Thus, the capacity upper
bound of \thmref{thm:SWOTCapUpper} matches the capacity lower bound
of \thmref{thm:SWOTRateLower}, implying the following corollary.

\begin{corollary}
The $1$-out-of-$m$ sample-wise OT capacity for the uniform source
distribution and correlated randomness in the form of a BES/BEC$(p)$
is given by
\[
C_{OT,m} = \min((1-p),p/(m-1)).
\]
The {\bf SWOT} protocol achieves capacity.
\end{corollary}

Note that this capacity is maximized at $C = (1/m)$ for the erasure
probability $p^* = (m-1)/m$, where the ratio of erasures to
non-erasures matches the ratio of bits concealed to bits revealed.
The {\bf SWOT} protocol achieves capacity since it efficiently
utilizes the erasures and non-erasures in revealing and concealing
the appropriate bits.


\section{String OT with Disjoint Privacy} \label{sec:Boot}
The bootstrap OT ({\bf BOOT}) protocol addresses the problem of
string OT with the disjoint privacy condition
\eqnref{eqn:weakPrivacy}. The {\bf SWOT} protocol could also be
applied to this problem, yielding the achievable rate given in
\thmref{thm:SWOTRateLower} with the stronger sense of joint privacy
\eqnref{eqn:alicePriv}. However, the {\bf BOOT} protocol achieves
rates that are better by a factor up to $((m-1)/\lceil \log_2 m
\rceil)$ (when the probability of erasure $p \leq 1/2$) since it
provides only disjoint privacy.

\subsection{Bootstrap String Oblivious Transfer Protocol} \label{sec:protocolBootOT}
The {\bf BOOT} protocol for $1$-out-of-$m$ string OT is
parameterized by a finite sequence of $u$ integers, $s_1,\ldots,s_u
\in \{2,\ldots,m\}$, such that $\prod_{i=1}^u s_i \geq m$. The {\bf
BOOT} protocol leverages $u$ uses of the {\bf SWOT} protocol, where
the $i$-th usage is for $1$-out-of-$s_i$ OT. For $i = 1,\ldots,u$,
Alice generates $s_i$, independent $k$-bit Bernoulli-$(1/2)$ masking
strings $\{\tilde{Z}_{i,j}\}_{j=1}^{s_i}$. The basic idea is to
encode each one of Alice's strings with the XOR of a different
combination of $u$ of these masking strings, taking one from each
set $\{\tilde{Z}_{i,j}\}_{j=1}^{s_i}$ for $i = 1,\ldots,u$. Alice
first sends these encodings, denoted by $\tilde{C}_1, \ldots,
\tilde{C}_m$, to Bob over the discussion channel. Then, for Bob to
decode a particular string of Alice's, Alice and Bob perform $u$
oblivious transfers where in the $i$-th OT Bob chooses from
$\{\tilde{Z}_{i,j}\}_{j=1}^{s_i}$ the masking string that is part of
the combination masking the string of Alice's that he wants. The
method in which each string of Alice is assigned a unique
combination of masking strings can be visualized by a tree
structure.

\begin{figure}
\centering
\includegraphics[width=2.0in]{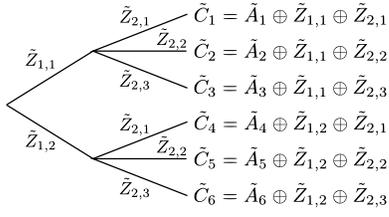}
\caption{The encoding tree structure for the {\bf BOOT} protocol for
$1$-out-of-$6$ string OT, with parameters $u = 2$, $s_1 = 2$ and
$s_2 = 3$.} \label{fig:bootstrap}
\end{figure}

The encoding tree structure for the example of $1$-out-of-$6$ string
OT via the {\bf BOOT} protocol with parameters $u = 2$, $s_1 = 2$
and $s_2 = 3$ is illustrated in \figref{fig:bootstrap}. In this
example, if Bob wishes to obtain $\tilde{A}_3$, he would select
$\tilde{Z}_{1,1}$ in first round of OT and then select
$\tilde{Z}_{2,3}$ in second round of OT, allowing him to reconstruct
$\tilde{A}_3$ via $\tilde{C}_3 \oplus \tilde{Z}_{1,1} \oplus
\tilde{Z}_{2,3}$.

\subsection{Analysis and Achievable Rates} \label{sec:mainWeakOT}
The perfect privacy of the {\bf SWOT} protocol guarantees that Bob
only learns the particular combination of masking strings
$\tilde{Z}_{i,j}$ that he selected, however, by the structure of the
encoding, its possible for Bob to learn some information about
Alice's strings beyond just the knowledge of $\tilde{A}_B$. However,
Bob will not be able to determine the specific value of any
particular string $\tilde{A}_i$ for $i \neq B$. Consider the example
of $1$-out-of-$6$ string OT illustrated in \figref{fig:bootstrap}
for $B = 3$. Since Bob learns $\tilde{Z}_{1,1}$ and
$\tilde{Z}_{2,3}$, he can also determine certain joint functions of
Alice's strings such as $\tilde{A}_1 \oplus \tilde{A}_4 \oplus
\tilde{A}_6$, which can be found from $\tilde{C}_1 \oplus
\tilde{C}_4 \oplus \tilde{C}_6 \oplus \tilde{Z}_{1,1} \oplus
\tilde{Z}_{2,3}$. Note that however, Bob cannot reduce any of the
equations further to determine the value of any $\tilde{A}_i$ for $i
\neq 3$. The proof for the general situation is omitted due to space
limitations.

The correctness of the protocol follows if each of the $u$ usages of
the {\bf SWOT} protocol is correct, which happens if there is enough
erasures and non-erasures in BES/BEC samples in each usage. Note
that instead, all of the BES/BEC samples can be taken at the
beginning, with the erasures and non-erasures being allocated to
multiple usages if they are sufficient, which will happen with high
probability for $n$ sufficiently large provided that $k/n$ is
slightly less than the achievable rate determined by the following
rate analysis.

For each round $i = 1,\ldots,u$, the {\bf SWOT} protocol for
$1$-out-of-$s_i$ OT of $k$-bit strings requires asymptotically $n_i
= k/R_{OT,s_i}$ samples of the BES/BEC. The total number of samples
of BES/BEC needed is $n = \sum_{i=1}^u n_i = \sum_{i=1}^u
k/R_{OT,s_i}$. Thus, the asymptotic rate achieved by this protocol
is given by the following theorem.

\begin{theorem} \label{thm:bootstrapRate}
Let $s_1,\ldots,s_u \in \{2,\ldots,m\}$ be a finite sequence of
integers such that $\prod_{i=1}^u s_i \geq m$. Then, the {\bf BOOT}
protocol with parameters $(s_1,\ldots,s_u)$ for a BES/BEC$(p)$
obtains perfect disjoint privacy and achieves the string OT rate
\[
\widetilde{R}_{OT,m} = \left( \sum_{i=1}^u \frac{1}{R_{OT,s_i}}
\right)^{-1},
\]
where $R_{OT,s} = \min((1-p),p/(s-1))$ is the achievable sample-wise
OT rate for the BES/BEC$(p)$ from \thmref{thm:SWOTRateLower}. Hence,
the string OT capacity (with disjoint privacy) is bounded below by
\[
\widetilde{C}_{OT,m} \geq \max_{u,s_1,\ldots,s_u} \left(
\sum_{i=1}^u \frac{1}{R_{OT,s_i}} \right)^{-1},
\]
where the maximization is taken over the set of finite sequences of
integers $s_1,\ldots,s_u \in \{2,\ldots,m\}$ such that
$\prod_{i=1}^u s_i \geq m$.
\end{theorem}

The {\bf SWOT} protocol is most efficient for erasure probability
$p^* = (m-1)/m$, since it needs a large proportion of erasures to
fully conceal the $k(m-1)$ bits that Bob did not select. The best
erasure probability for the {\bf BOOT} protocol, however, is
variable and depends on the choice of parameters. For example,
setting $u = \lceil \log_2 m \rceil$ and $s_i = 2$ for all $i$,
yields the achievable rate
\[
\widetilde{R}_{OT,m} = \frac{R_{OT,2}}{\lceil \log_2 m \rceil} =
\frac{\min(p,(1-p))}{\lceil \log_2 m \rceil},
\]
which is maximized at $\widetilde{R}_{OT,m} = 1/(2\lceil \log_2 m
\rceil)$ for $p = (1/2)$. Comparing this to the achievable
sample-wise OT rate for $p = (1/2)$, $R_{OT,m} = 1/(2(m-1))$ reveals
an improvement in rate by a factor of $((m-1)/(\lceil \log_2 m
\rceil))$. The {\bf BOOT} protocol achieves higher rates since it
effectively recycles the erasures to conceal more bits. Some
information is leaked since the erasures are being recycled,
however, only joint functions of the strings (specifically the
exclusive-or of multiple strings) are revealed while maintaining the
disjoint privacy. Note that for the parameters $u = 1$ and $s_1 =
m$, the {\bf BOOT} protocol achieves the same rate as the {\bf SWOT}
protocol. Thus, the {\bf BOOT} protocol can achieve any rate
achieved by the {\bf SWOT} protocol.

\begin{figure}
\centering
\includegraphics[width=2.8in]{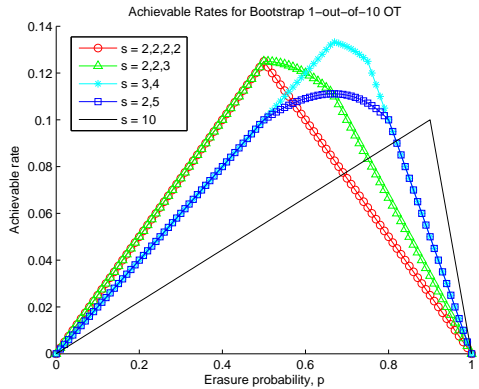}
\caption{Achievable rates of the {\bf BOOT} protocol for
$1$-out-of-$10$ string OT as a function of erasure probability $p$
of the BES/BEC$(p)$. Each curve represents the achievable rates for
a different set of parameters of the {\bf BOOT} protocol.}
\label{fig:bootrates}
\end{figure}

\figref{fig:bootrates} illustrates the achievable rates of the {\bf
BOOT} protocol, with $m=10$, as a function of erasure probability
$p$ of the BES/BEC$(p)$, for various sets of parameters. Note that
in different ranges, different sets of parameters are best. The {\bf
BOOT} protocol for parameters $\{s_1 = 10\}$ (giving the performance
of the {\bf SWOT} protocol) is best only in the range of erasure
probability close to $p = (m-1)/m$ and above.


\section{Secure Function Computation}

The general secure function computation ({\bf GSFC}) protocol
leverages the {\bf SWOT} protocol. It uses two oblivious transfers,
where the first is from Alice to Bob and the second is {\em from Bob
to Alice}, reversing the roles. Since the {\bf SWOT} protocol uses a
BES/BEC in the direction of the transfer, the {\bf GSFC} protocol
uses a BES/BEC available in both directions. The rate is determined
as the ratio of function samples $k$ to the total number of BES/BEC
samples used in both directions.

\subsection{General SFC Protocol} \label{sec:protocolFuncComp}
This protocol is applicable to any general sources and functions.
Without loss of generality, let the finite source alphabets be given
by $\mathcal{A} = \{1,\ldots,m_A\}$ and $\mathcal{B} =
\{1,\ldots,m_B\}$, and the ranges of the functions $f$ and $g$ be
$\mathcal{R}_f = \{0,1\}^{h_A}$ and $\mathcal{R}_g = \{0,1\}^{h_B}$
respectively.

We outline the {\bf GSFC} protocol with the following steps. For Bob
to compute $G^k$, Alice generates $m_B$, $kh_B$-bit strings, for $i
= 1, \ldots, m_B$, $\widetilde{A}'_i = (g(A_1,i), \ldots,
g(A_k,i))$. Bob expands his $k$ source samples $B^k$ to a vector of
length $kh_B$, where each element of $B^k$ is repeated $h_B$ times
to produce the samples $B'^{kh_B}$. Alice and Bob then use the {\bf
SWOT} protocol to perform $1$-out-of-$m_B$ OT for $kh_B$-bit strings
with $\{\widetilde{A}'_i\}_{i=1}^{m_B}$ as Alice's strings, and
Bob's selections vector as $B'^{kh_B}$. The result of this OT gives
Bob $(g(A_1,B_1), \ldots, g(A_k,B_k))$. Similarly, for Alice to
compute $F^k$, Alice and Bob reverse roles and perform
$1$-out-of-$m_A$ OT for $kh_B$-bit strings from Bob to Alice.

\subsection{Analysis and Achievable Rates}
The perfect privacy, negligible probability of error, and
distribution-free properties of {\bf SWOT} protocol imply the same
properties in this secure function computation protocol. The
$1$-out-of-$m_B$ OT for $kh_B$-bit strings via the {\bf SWOT}
protocol asymptotically requires $n_1 = kh_B/R_{OT,m_B}$ samples of
a BES/BEC from Alice to Bob, and likewise the other OT requires $n_2
= kh_A/R_{OT,m_A}$ samples of a BES/BEC from Bob to Alice, yielding
the following theorem.

\begin{theorem} \label{thm:funcCompRate}
Let $m_A = |\mathcal{A}|$, $m_B = |\mathcal{B}|$, $h_A = \lceil
\log_2 |\mathcal{R}_f| \rceil$, and $h_B = \lceil \log_2
|\mathcal{R}_g| \rceil$. Then, the {\bf GSFC} protocol, utilizing
correlated randomness in the form a BES/BEC$(p)$ available in both
directions, is perfectly-private, distribution-free and achieves the
function computation rate
\[
R = \left(\frac{h_B}{R_{OT,m_A}} +
\frac{h_A}{R_{OT,m_B}}\right)^{-1},
\]
where $R_{OT,m} = \min((1-p),p/(m-1))$ is the achievable sample-wise
OT rate from \thmref{thm:SWOTRateLower}. The function computation
capacity is bounded below by $C \geq R$.
\end{theorem}

The {\bf GSFC} protocol is general, but not optimal since it does
not exploit any source correlation or functional structure. Note
that only one usage of the {\bf SWOT} is necessary if one of the
functions $f, g$ is a function of (or the same as) the other
function.


\bibliographystyle{IEEEtran}
\bibliography{../LaTeX/biblio}

\end{document}